\documentclass[12pt,preprint]{aastex}
\usepackage{url}
\usepackage{multirow}
\usepackage[symbol*]{footmisc}

\newcommand{\sat}[1]{\it\uppercase{#1}\rm}
\newcommand{\fig}[1]{Figure~\ref{#1}}
\newcommand{\tbl}[1]{Table~\ref{#1}}
\newcommand{\speed}[1]{#1 km s${}^{-1}$}
\newcommand{\aspeed}[1]{$\sim$#1 km s${}^{-1}$}
\newcommand{\rsun}[1]{${#1}\,R_\sun$}

\shorttitle{} %

\shortauthors{Liu, Liu \& Wang}

\begin{document}

\title{Gradual Inflation of Active-Region Coronal Arcades Building up to Coronal Mass Ejections}

\author{Rui Liu, Chang Liu, Sung-Hong Park, \& Haimin Wang}
\affil{Space Weather Research Laboratory, Center for Solar-Terrestrial Research, NJIT, Newark, NJ
07102; rui.liu@njit.edu}

\begin{abstract}
The pre-CME structure is of great importance to understanding the origin of CMEs, which, however,
has been largely unknown for CMEs originating from active regions. In this paper, we investigate
this issue using the wavelet-enhanced EUV Imaging Telescope observations combined with the Large
Angle and Spectrometric Coronagraph, Michelson Doppler Imager, and \sat{goes} soft X-ray
observations. Selected for studying are 16 active-region coronal arcades whose gradual inflation
lead up to CMEs. 12 of them clearly build upon post-eruptive arcades resulting from a preceding
eruption; the rest 4 are located high in the corona in the first place and/or have existed for
days. The observed inflation sustains for $8.7\pm4.1$ h, with the arcade rising from
$1.15\pm0.06\,R_\sun$ to $1.36\pm0.07\,R_\sun$ within the EIT field of view (FOV). The rising speed
is less than \speed{5} for most of the time. Only at the end of this quasi-static stage, it
increases to tens of kilometers per second over tens of minutes. The arcade then erupts out of the
EIT FOV as a CME with similar morphology. This pre-CME structure is apparently unaffected by the
flares occurring during its quasi-static inflation phase, but is closely coupled with the flare
occurring during its acceleration phase. For four events that are observed on the disk, it is found
that the gradual inflation of the arcade is accompanied by significant helicity injection from
photosphere. In particular, a swirling structure, which is reminiscent of a magnetic flux rope, was
observed in one of the arcades over 4 h prior to the subsequent CME, and the growth of the arcade
is associated with the injection of helicity of opposite sign into the active region via flux
emergence. We propose a four-phase evolution paradigm for the observed CMEs, i.e., a quasi-static
inflation phase which corresponds to the buildup of magnetic free energy in the corona, followed by
the frequently observed three-phase paradigm, including an initial phase, an acceleration phase and
a gradual phase.

\end{abstract}


\section{Introduction}

Coronal mass ejections (CMEs) are complex, large-scale magnetized plasma expelled from the Sun,
often intertwined with flares and prominence eruptions.  A classic, white-light CME displays a
three-part structure: a bright expanding loop, followed by a dark void (cavity) embedded with a
bright core of dense prominence material \citep[e.g.,][]{ih86}. Typically, it originates from a
helmet streamer that has been slowly rising or swelling for days prior to the eruption
\citep[see][and references therein]{gibson06}. The helmet streamer also possesses an equivalent
three-part structure, corresponding to a filament channel observed on the solar disk. During the
eruption, the streamer deforms into the CME frontal structure; the prominence and the cavity
underlying the streamer become the bright core and the cavity of the CME, respectively. The
prominence is often long-lived and can be tracked on the solar disk for several days, and its
activation phase with a gradual rising motion at tens of kilometers per second may precede the CME
eruption by tens of minutes or even hours \citep[e.g.,][]{sss99}. Great importance has been rightly
attached to these quiescent structures, which provide important clues to the nature of coronal
magnetic fields prior to CMEs.

It has long been proposed that there are two dynamical types of CMEs \citep{gosling76, mf83,
sheeley99}: 1) fast CMEs, which are accelerated impulsively at the low corona and decelerated in
the coronagraph FOV, preferentially associated with solar flares; and 2) slow CMEs, which are
accelerated gradually in the coronagraph FOV over a large height range, preferentially associated
with prominence eruptions. However, \citet{vsr05} found that non-flare CMEs show characteristics
similar to CMEs associated with flares of soft X-ray class B and C, which is indicative of a
``continuum'' of events rather than supporting the existence of two distinct classes. A more
compelling argument for a one-class continuum view is based on a statistical study of 4315 CMEs by
\citet{yurchy05}, who showed that the speed distributions for both accelerating and decelerating
events can be fitted with a single log-normal distribution to a good approximation.

In contrast to CMEs associated with prominence eruptions, not just the pre-CME coronal structure
but its physical connection with the subsequent CME is poorly identified for CMEs associated with
flares. To establish such a connection may help understand the conditions leading up to the
catastrophic release of the magnetic free energy. With the helmet streamer in mind, we assume that
a pre-CME structure in general should have a closed magnetic configuration that evolves with a
quiescent energy buildup phase, followed by a catastrophic eruption into a CME which bears
morphological similarity with the pre-CME structure.

It has been shown that active regions that exhibit forward- or reverse-S shapes in soft X-rays,
also termed sigmoids, have a greater tendency to erupt \citep{chm99}, viz., sigmoids often brighten
prior to or during the flare impulsive phase, and then transform into arcades or cusped loops as
the eruption progresses \citep{sterling00, moore01, pevtsov02, liu07}. The sigmoidal shape is
indicative of the presence of twisted magnetic flux. Hence its formation may be driven by the
helicity injection via shearing of photospheric fields or direct emergence of twisted flux. But it
is unknown in observation which part of the sigmoidal structure has been expelled as part of the
CME, despite that a topological reconfiguration has occurred presumably due to magnetic
reconnection.

On the other hand, although coronal loops, especially post-flare loops, are the most prominent
closed-field structures in active regions, they usually take a relaxed, bipolar shape, hence are
not expected to bear much free energy that necessitates the birth of flares or CMEs. During solar
flares, the formation of post-flare loops is associated with the motion of flare ribbons, which are
essentially composed of the loop footpoints. The whole arcade of loops is also referred to as the
post-eruptive arcade. It is often observed with filters of decreasing temperatures as time
progresses, indicative of progressive cooling \citep[e.g.,][]{aa01}. \citet{tbc04} studied 236
post-eruptive arcades in EIT 195 {\AA} from 1997 to the end of 2002, whose average EUV emission
life-time ranged from 2 to 20 h, with an average of 7 h. \citet{lz09} studied the early evolution
of \sat{trace} post-flare loops resulting from 190 M- and X-class flares from May 1998 to December
2006. None of these authors reported that any post-eruptive arcade become eruptive again during its
lifetime.

In a recent study of coronal loops, to our surprise, we noticed a number of events showing the
gradual growth of an EUV post-eruptive arcade leading up to a CME, with the CME front bearing
morphological similarities to the arcade. A similar behavior is found for some active-region loops
that are located high in the corona in the first place and/or have existed for days prior to the
gradual growth, referred to overlying arcades hereafter. In this paper, we study these swelling
coronal arcades and explore whether they represent a distinct pre-CME structure. The rest of the
paper is organized as follows: we present the observations and data analysis in Section 2. After an
overview of the observations (\S2.1), a few selected events are studied in more detail.
Particularly, the gradual growth and the subsequent eruption of post-eruptive arcades are presented
in \S2.2, and that of overlying arcades in \S2.3. We discuss in Section 3 the implications for CME
physics.

\section{Observation}

\subsection{Overview}
\begin{deluxetable}{cccccccccccc}
\tablecolumns{12} %
\tabletypesize{\scriptsize} %
\tablewidth{0pt} %
\tablecaption{List of Events \label{list}}%
\tablehead{\multirow{3}{*}{No.} & \multicolumn{6}{c}{Arcade} & \multicolumn{2}{c}{Flare\tablenotemark{c}} & \multicolumn{3}{c}{CME} \\
& \multirow{2}{*}{$t_i$} & \colhead{$\Delta t$\tablenotemark{a}} & \colhead{$r_i$} &
\colhead{$r_f$} & \colhead{$v_q$\tablenotemark{b}} & \multirow{2}{*}{Type} &
\multirow{2}{*}{Location}
& \multirow{2}{*}{GOES} & \colhead{PA/AW\tablenotemark{d}} & \colhead{$v$\tablenotemark{e}} & \colhead{$a$\tablenotemark{f}} \\
& & (hr) & (R${}_\sun$) & (R${}_\sun$) & (km/s) & & & & (deg) & (km/s) & (m s${}^{-2}$) }
\startdata %
1  & 1998-11-23 22:33 & 3.7  & 1.14 & 1.27 & 3.6  & PEA & S29W90    & X1.0 & 220/Halo & 1798 & -12.5\\
2  & 2001-02-02 12:00 & 6.5  & 1.27 & 1.38 & 1.7  & OA & N32E74    & C3.3 & 67/153   & 845  & 7.5 \\
3  & 2001-04-15 00:00 & 8.8  & 1.20 & 1.38 & 1.5  & OA & S20W85    & C6.7 & 330/64   & 301  & -6.9* \\
4  & 2002-03-23 23:12 & 12.0 & 1.19 & 1.48 & 4.7  & PEA & Occulted? & N/A  & 247/62   & 375  & 0.0* \\
5  & 2003-06-15 19:25 & 4.2  & 1.14 & 1.28 & 5.8  & PEA & S07E80    & X1.3 & 122/Halo & 2053 & -0.9 \\
6  & 2005-01-15 12:12 & 10.4 & 1.11 & 1.35 & 2.6  & PEA & N15W05    & X2.6 & 323/Halo & 2861 & -127.4* \\
7  & 2005-01-20 00:00 & 6.6  & 1.22 & 1.41 & 2.3  & PEA & N14W61    & X7.1 & 291/Halo & 882  & 16.0* \\
8  & 2005-07-13 05:00 & 9.0  & 1.04 & 1.32 & 3.7  & PEA & N11W90    & M5.0 & 300/Halo & 1423 & -14.1 \\
9  & 2005-07-13 20:36 & 8.2  & 1.09 & 1.38 & 5.4  & PEA & N10W90    & M1.0 & 282/60   & 541  & -3.4 \\
10 & 2005-07-14 05:00 & 5.4  & 1.19 & 1.41 & 5.1  & PEA & N11W90    & X1.2 & 281/Halo & 2115 & 198.0*\\
11 & 2005-07-14 22:24 & 11.6 & 1.12 & 1.36 & 3.2  & PEA & Occulted? & N/A  & 282/62   & 629  & 39.1 \\
12 & 2005-08-22 04:00 & 12.8 & 1.10 & 1.30 & 2.0  & PEA & S13W65    & M5.6 & 251/Halo & 2378 & 108.0* \\
13 & 2005-08-23 02:00 & 12.4 & 1.13 & 1.30 & 1.9  & PEA & S14W90    & M2.7 & 220/Halo & 1929 & 44.2 \\
14 & 2005-08-23 13:36 & 18.6 & 1.15 & 1.47 & 2.3  & PEA & Occulted? & N/A  & 75/24    & 726  & 98.5* \\
15 & 2006-05-01 10:14 & 4.8  & 1.12 & 1.23 & 2.6  & OA & S12W27    & C1.0 & 180/Halo & 487  & 3.0* \\
16 & 2006-07-04 14:48 & 4.6  & 1.22 & 1.36 & 4.3  & OA & S13W14    & C1.4 & 199/102  & 308  & 1.6 \\
\hline %
$\bar{x}$ & & 8.7 & 1.15 & 1.36 & 3.3 & & & & & 1228 & 21.9 \\
$\sigma$  & & 4.1 & 0.06 & 0.07 & 1.4 & & & & & 842  & 69.9 \\ \hline

\enddata
\tablecomments{} %
\tablenotetext{a}{Time duration during which the apex height of the swelling arcade is measured
within the EIT FOV; $r_i$ and $r_f$ are measured at $t_i$ and $t_i+\Delta t$, respectively. }

\tablenotetext{b}{Speed obtained by linear fit of the height-time profile at the quasi-static
stage.}

\tablenotetext{c}{For the event in which the corresponding active region was behind the limb, the
associated flare might be occulted, therefore not being detected by \sat{goes} (``N/A'').}

\tablenotetext{d}{Position Angle/Angular Width, which are given by the LASCO CME Catalog except
that for halo CMEs the position angle refers to that of the fiducial adopted.} %

\tablenotetext{e}{~Linear speed given by the LASCO CME Catalog.} %

\tablenotetext{f}{Acceleration given by the LASCO CME Catalog. The * symbol indicates that
acceleration is uncertain due to either poor height measurement or a small number of height-time
measurements. }
\end{deluxetable}

\begin{figure}\epsscale{1}
\plotone{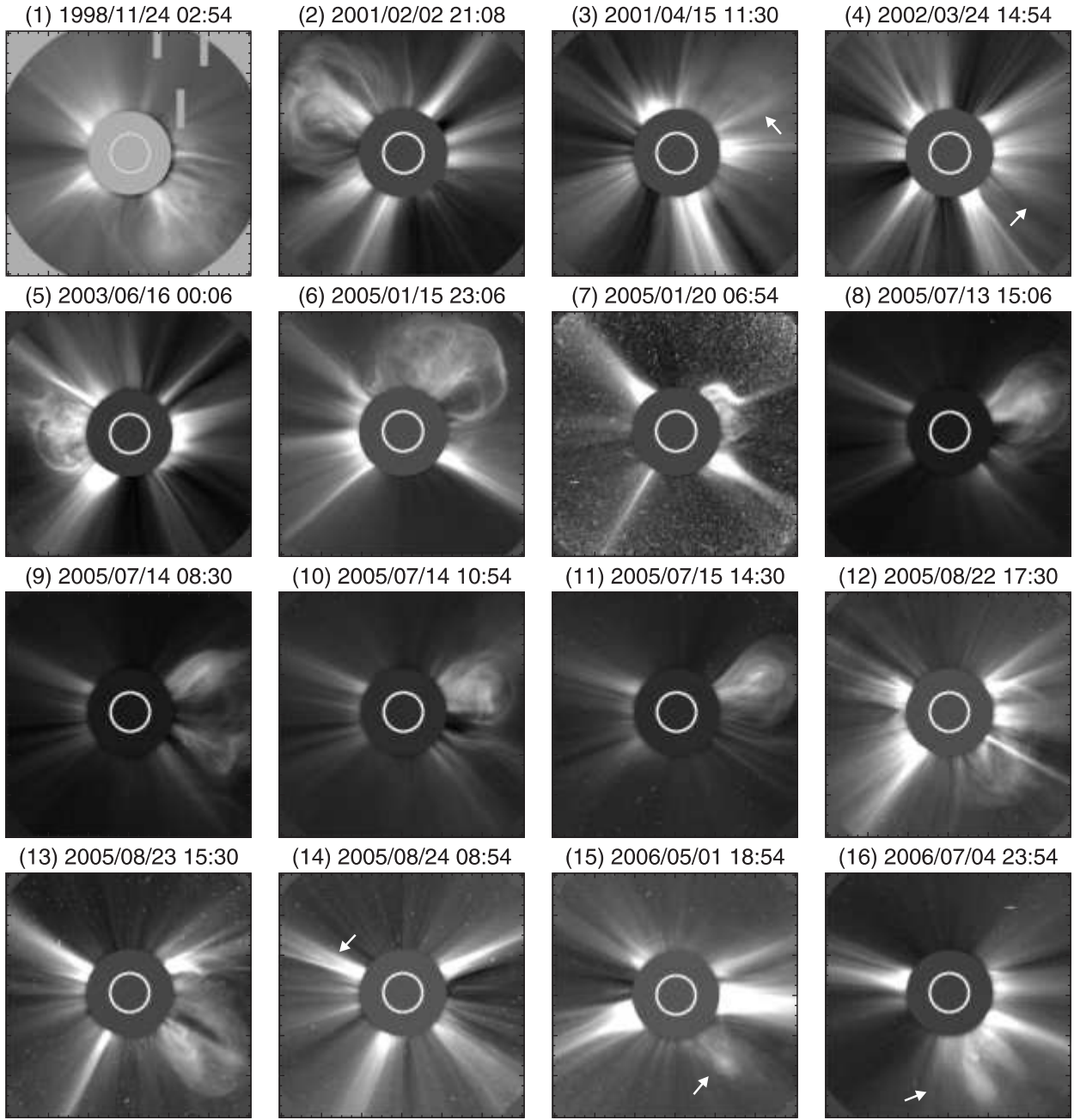} \caption{\sat{soho} LASCO C2 images showing the CMEs originating from inflating
active-region arcades. CMEs that are not quite discernible are marked by arrows. The animation of
corresponding C2 difference images are available at the LASCO CME Catalogue. The event No. given in
\tbl{list} is indicated in the image titles. \label{c2}}
\end{figure}

In this study, 16 events are selected as listed below in \tbl{list}, which all feature clearly the
gradual expansion of a coronal arcade leading up to a CME. The event selection is based on a survey
of the \sat{soho} EIT 195 {\AA} data from January 1997 to October 2006, which is greatly
facilitated by the ``wavelet'' movies
\citep{svh08}\footnote{\url{http://lasco-www.nrl.navy.mil/index.php?p=content/wavelet}}. These
events could have been easily missed in the raw data, but are clearly seen in the wavelet-enhanced
images. They are composed of two groups of events, 1) post-eruptive arcades (PEAs; 12 events); and
2) overlying arcades (OAs; 4 events). None of the active regions of interest are of a sigmoidal
shape in \it{Yohokoh} SXT or \sat{goes} SXI soft X-ray images prior to the eruption. Instead, soft
X-ray data often show a diffuse arcade similar in morphology to the one observed in EUV during the
gradual inflation stage, and a bright flare loop in the wake of the eruption. A note of caveat has
to be made that in this study we have left out those events in which one is unable to track the
same loop, e.g., a) the arcade expands beyond the EIT FOV during the early stage, which partly
accounts for the scarcity of the OA events; b) higher loops of the arcade become too dim to be seen
beyond a certain height. 

To obtain the hight-time profile of the swelling arcades, we choose a reference point on the solar
surface, draw a fiducial along the growing direction of the arcade, and record the highest point
that the fiducial intersects the arcade, assuming that the arcade is oriented vertically. The
reference point as well as the fiducial is made to rotate with the sun. If the arcade grows in the
radial direction, the ``true'' height of the arcade (in $R_\sun$) is obtained by dividing its
projected height with respect to the Sun center by the projected distance of the reference point
from the Sun center. However, if the arcade is located on the limb, or its growth obviously
deviates from the radial direction, only the projected distance from the arcade apex to the
reference point is recorded. The distance after adding by the solar radius is then regarded as the
heliocentric distance. The height of the resultant CME with time in the LASCO FOV is readily
available from the LASCO CME Catalogue\footnote{\url{http://cdaw.gsfc.nasa.gov/CME_list/}}. Type II
radio emission, if existent, can give some idea of the evolution of the CME in the gap between the
FOV of EIT and that of LASCO C2 (from \rsun{1.5} to \rsun{2.2}), since it is generally interpreted
as plasma emission near the local electron plasma frequency due to electrons accelerated by shock
waves. The height of the supposed shock-front is obtained by examining the slowly drifting bands of
emission in the radio dynamical spectra. However, it is often hard to tell not just whether the
shock nose or the shock flank is observed, but whether the shock is flare-driven or CME-driven
\citep[c.f.,][]{mr04}

The results of the measurement are listed in \tbl{list}. On average, the arcades grow from
$1.15\pm0.06$ R${}_\sun$ to $1.36\pm0.07$ R${}_\sun$ during $8.7\pm4.1$ hrs within the EIT FOV. The
speed during the linear, quasi-static stage is $3.3\pm1.4$ km s${}^{-1}$. However, the resultant
CMEs observed by \sat{soho} LASCO are quite diverse, with a speed of $1228\pm842$ km s${}^{-1}$
from the linear fit, and with an acceleration of $21.9\pm69.9$ m s${}^{-2}$ from the parabolic fit.
As shown in \fig{c2}, one can see that the morphologies of the CMEs are also quite diverse. Some of
them, especially Events 8, 9, 10, 11 and 13, are reminiscent of the so-called ``flux-rope CMEs''
\citep[e.g.,][]{dere99}, and some of them, especially Events 5 and 6, give an impression of an
arcade of loops. Only one narrow jet-like CME (Event 14) is observed for this set of events. It is
also interesting that 9 of the 16 events occurred in 2005. In particular, Events 6 and 7 occurred
in AR 10720, Events 8--10 in AR 10786, and Events 12 and 13 in AR 10798. This implies that such
events may be specific to active region properties.

Despite the limited number of events, interestingly, CMEs building upon PEAs are obviously more
energetic than those evolving from OAs, with the former propagating at a much faster speed
($1476\pm826$ km s${}^{-1}$) than the latter ($485\pm255$ km s${}^{-1}$) within the LASCO FOV.
Moreover, 8 of the 12 PEA events results in Halo CMEs while only 1 slow Halo CME (Event 15) for the
4 OA events. The PEA events are also associated with larger flares (M- and X-class) than the OA
events (C-class). A question remains as to whether PEAs are distinct from OAs. Morphologically
speaking, an eruption building upon a PEA often involves a series of loops spreading along the
neutral line, while an OA often consists of a bundle of loops sharing compact footpoint regions.
The difference may pertain to the energetics. But there is no clear distinction between PEAs and
OAs in terms of the dynamical evolution, and it is hard to tell whether an OA actually evolves from
a PEA.

Since coronal arcades are preferentially seen against dark background, most of the events are
observed above the limb, except Events 6, 7, 12, 15, and 16. In Event 7, however, the east-west
oriented arcade is located at about 60 degree to the west of the central meridian, and can only be
partially seen. Hence, we will concentrate on Events 6, 12, 15 and 16 in the following subsections.
Events 6 and 12 (\S2.2) are PEA events, while Events 15 and 16 (\S2.3) are of the OA type. Although
Event 13 is also a limb event, the arcade is oriented north-south, and most importantly it resulted
directly from the eruption in Event 12. Hence we will study Events 12 and 13 together (\S2.2.2).

For the four disk events, we utilize MDI magnetograms to estimate the helicity accumulation in the
relevant active region. We first apply the local cross-correlation tracking (LCT) method
\citep{ns88} to estimate the change rate of relative magnetic helicity in an open volume through a
boundary surface $S$ \citep{cmp04, cj05}, viz.,
\[
\left(\frac{dH_{m}}{dt}\right)_\mathrm{LCT} = -2\int_{S}B_n({\bf{v}}_\mathrm{LCT} \cdot
{\bf{A}}_p)\,dS.
\]
where $B_n$ is the magnetic field component normal to the surface $S$, ${\bf v}_\mathrm{LCT}$ is
the apparent horizontal velocity field component determined by the LCT technique, and ${\bf A}_p$
is the vector potential of the potential field under the Coulomb gauge, viz., $ {\bf\hat{n}} \cdot
\nabla \times {\bf{A}}_{p} = B_{n}$, $\nabla \cdot {\bf{A}}_{p} = 0 $, and ${\bf{A}}_{p} \cdot
{\bf\hat{n}} = 0$. $B_n$ can be estimated from the MDI line-of-sight field, $B_l$, viz., $B_l= B_n
\cos\psi$, where $\psi$ is the heliocentric angle. ${\bf A}_p$ is then calculated from $B_n$ by
using the fast Fourier transform method in a usual fashion. We perform LCT for all pixels with an
absolute flux density greater than 5 G, but only those with cross correlation above 0.9 are used to
obtain ${\bf v}_\mathrm{LCT}$ \citep[for more details, see][]{park08}. The integration is carried
out over the entire area of the target active region. After the helicity change rate is determined
as a function of time, we integrate it with respect to time to obtain the amount of helicity
accumulation,
\[
\Delta H_{m} = \int_{t_0} ^{t} \left(\frac{dH_{m}}{dt}\right)_\mathrm{LCT} dt,
\]
where $t_0$ and $t$ are the start and end time of the helicity accumulation, respectively.

To estimate the uncertainty of helicity corresponding to the MDI measurement error ($\sim20$ G), we
carry out a Monte Carlo simulation by adding random noise to MDI magnetograms, i.e.,
$B_{l,\,\mathrm{sim}}=B_l+20\times r$, where $r$ is a pseudo-random number generated by the IDL
procedure \texttt{RANDOMN}. A series of helicity injection rates can be calculated based on these
simulated magnetograms. After repeating the above procedure for 10 times, we take the resultant
standard deviation as the uncertainty of the helicity injection rate. The uncertainty of the
helicity accumulation is then calculated following the rules of the propagation of uncertainty.

For all of the four disk events, consistently, there is significant helicity injection into the
corona prior to the flare associated with the eruption of the arcade. Details will be presented
below.

\subsection{Eruption of Post-Eruptive Arcades}

\subsubsection{Post-Eruptive Arcade on 2005 January 15}

\begin{figure}\epsscale{1}
\plotone{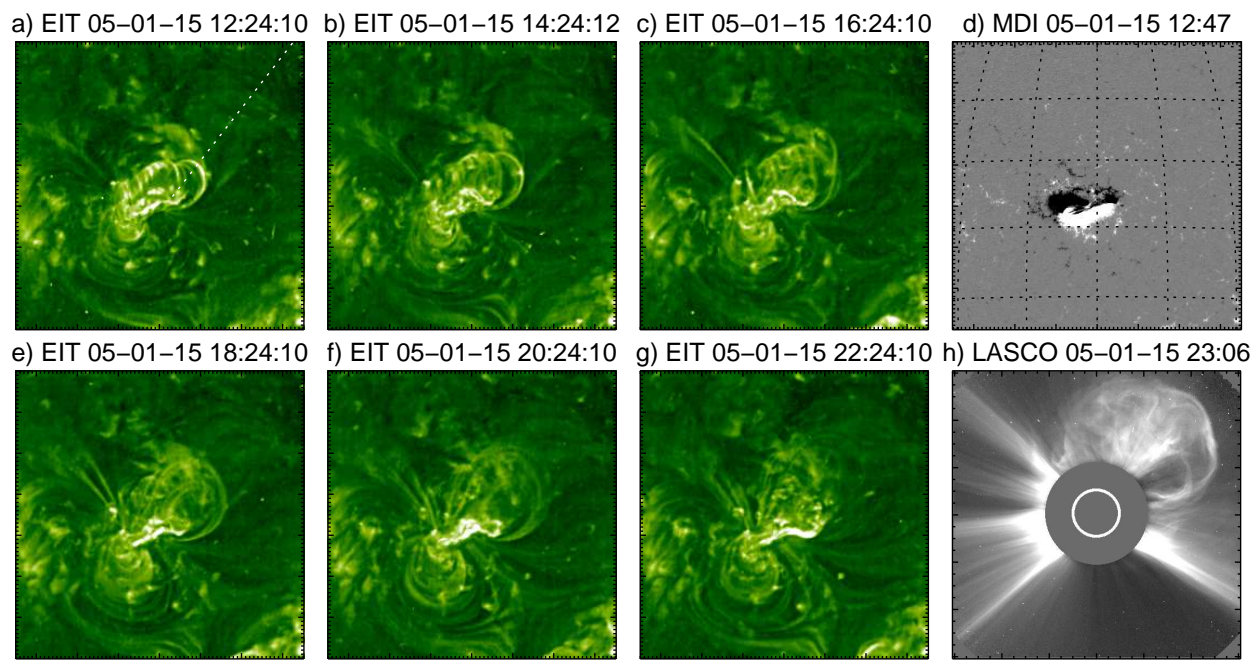} \caption{Evolution of the PEA observed on 2005 January 15. In frame (a) a
fiducial is drawn along the growing direction of the arcade to measure the height of the arcade
(see \S2.1 for details). The field of view in Panels (a--g) is 700 by 700 arcsecs, centering at
$(0'',\ 350'')$, with all images registered to the image in Panel (a). EIT images in this paper are
enhanced with a wavelet method based on \citet{sc03}. A video of EIT 195 {\AA} images is available
in the online edition of the Journal.\label{eit0501}}
\end{figure}

\begin{figure}\epsscale{0.7}
\plotone{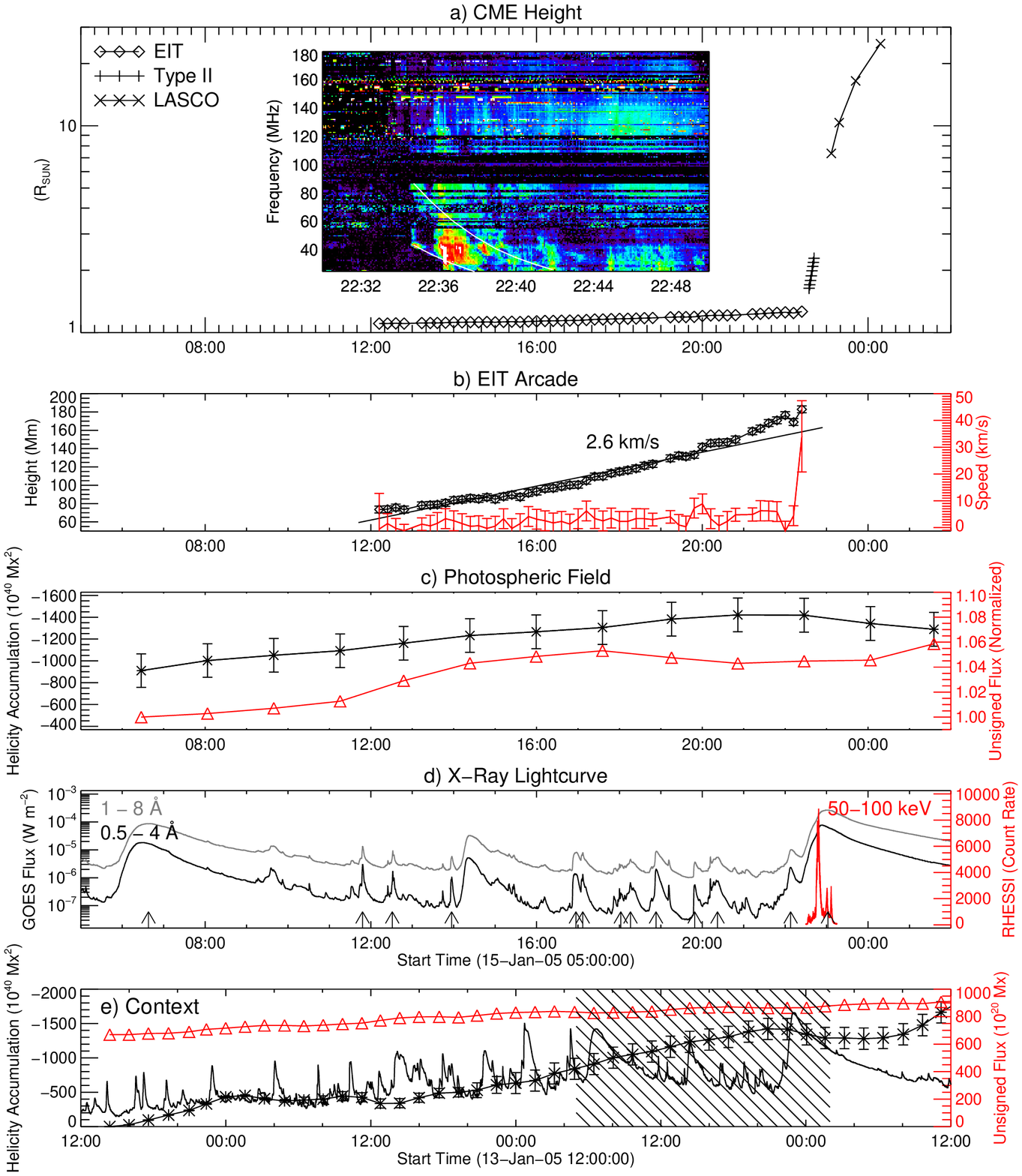} \caption{Height-time profile of the PEA and the resultant CME on 2005
January 15 in relation to the evolution of the photospheric magnetic field as well as X-ray
lightcurves. Panel (a) shows the height-time profiles of the EIT arcade, the shock front obtained
from Type II radio emission, and the CME front given by the LASCO CME catalogue, in solar radius
unit. The inset shows the radio dynamical spectra provided by the Radio Solar Telescope Network
(RSTN), with the two drifting bands of Type II emission denoted in dashed lines. In Panel (b), the
height-time profile of the EIT arcade is given in Mm unit, and the derived velocity-time profile is
displayed in red color and scaled by the y-axis on the right. Panel (c) shows the amount of
helicity accumulation (see \S2.1 for details) as well as the unsigned magnetic flux integrated over
the active region of interest. Panel (d) shows the \sat{goes} soft X-ray flux in 1-8 {\AA} (grey)
and 0.5--4 {\AA} (black), and the \sat{rhessi} count rate in 50--100 keV (red). For each flare of
\sat{goes}-class C and above occurring in AR 10720, we draw an arrow at the bottom to indicate its
soft X-ray peak. In Panel (e), the hatched box indicates the time duration covered by Panels
(a--d). \sat{goes} 1--8 {\AA} flux is displayed in an arbitrary unit. \label{ht0501}}
\end{figure}

The PEA on 2005 January 15 was located in the active region NOAA 10720 (\fig{eit0501}). The arcade
was produced by a Halo CME associated with a \sat{goes}-class M8.6 flare (E04N16), which peaked in
soft X-rays at 06:38 UT (\fig{ht0501}(d)). Its eruption about 16 hrs later resulted in a Halo CME
associated with an X2.6 flare with the peak in soft X-rays at 22:25 UT. The arcade formed as early
as about 06:24 UT, but we are only interested in its evolution beyond the end of the M8.6 flare at
about 12:00 UT, since the early rising of PEAs is largely attributed to the reconnection of
magnetic field lines at higher and higher altitudes in the corona \citep{pf02}. \fig{eit0501}(a)
displays a typical PEA which is composed of a series of bipolar coronal loops. The loop footpoints
constitute two bright, curved flare ribbons, which are parallel to each other and aligned along the
polarity inversion line of the line-of-sight photospheric field (\fig{eit0501}(d)). The gradual
inflation of the arcade is clearly demonstrated in \fig{eit0501}(a--c), but as time progressed,
most loops got more and more dim. As of 18:24 UT (\fig{eit0501}(e)), only visible are the loops at
the western section of the original arcade, whose height-time evolution is measured along a
fiducial as indicated by the dotted line in \fig{eit0501}(a).

Our measurement starts from the EIT image at 12:12 UT, when one can start to track the loops of
interest frame by frame. Moreover, at that time the soft X-ray flux has decreased to the background
level (\fig{ht0501}(d)) so that the effect of reconnection can be reasonably ignored. One can see
that despite multiple flares occurring in the same active region (as indicated by arrows at the
bottom of \fig{ht0501}(d)), the group of loops grew quasi-statically in height, at a speed of
\aspeed{2.6}, from about 12:12 UT until 22:24 UT when the speed suddenly increased to \aspeed{50},
coincidence with the onset of the flare. The inflation process was associated with significant
injection of negative helicity but with minimal flux increase (\fig{ht0501}(c)). It is interesting
that the amount of helicity accumulation were apparently ``saturated'' before the X2.6 flare
\citep[\fig{ht0501}(e);][]{park08}. In the next available EIT image at 22:36 UT, the loops became
invisible, but coronal dimming can be seen in the remote region close to the east limb (not shown),
as a manifestation of the blasted bubble sweeping across the disk. Just prior to the speed jump,
the arcade had a lower height at 22:12 UT than at both 22:00 and 22:24 UT. Note that below the
arcade, a filament eruption was frustrated, associated with an M1.0 flare peaking in soft X-rays at
22:08 UT \citep{liu10}. The contraction of the arcade was probably due to the filament being pulled
back to the surface.

\subsubsection{Homologous Eruptions of Post-Eruptive Arcades on 2005 August 22 and 23}

\begin{figure}\epsscale{0.55}
\plotone{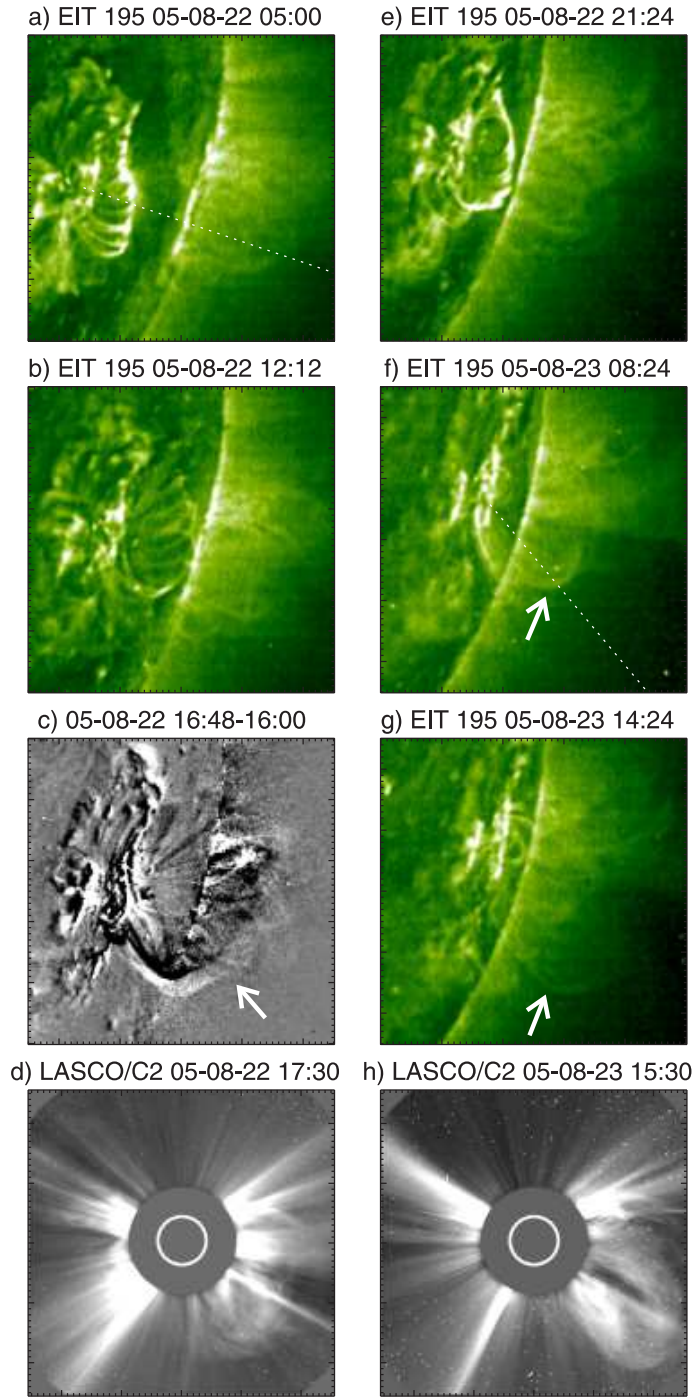} \caption{Eruptions of the two PEAs observed on 2005 August 22 (left column)
and 23 (right column) in the same active region AR 10798. The field of view in Panels (a--c) and
(e--g) is 500 by 500 arcsecs, centering at $(900'',\ -250'')$ and on $(950'',\ -300'')$,
respectively. A video of EIT 195 {\AA} images is available in the online edition of the Journal.
\label{eit0508}}
\end{figure}

\begin{figure}\epsscale{0.8}
\plotone{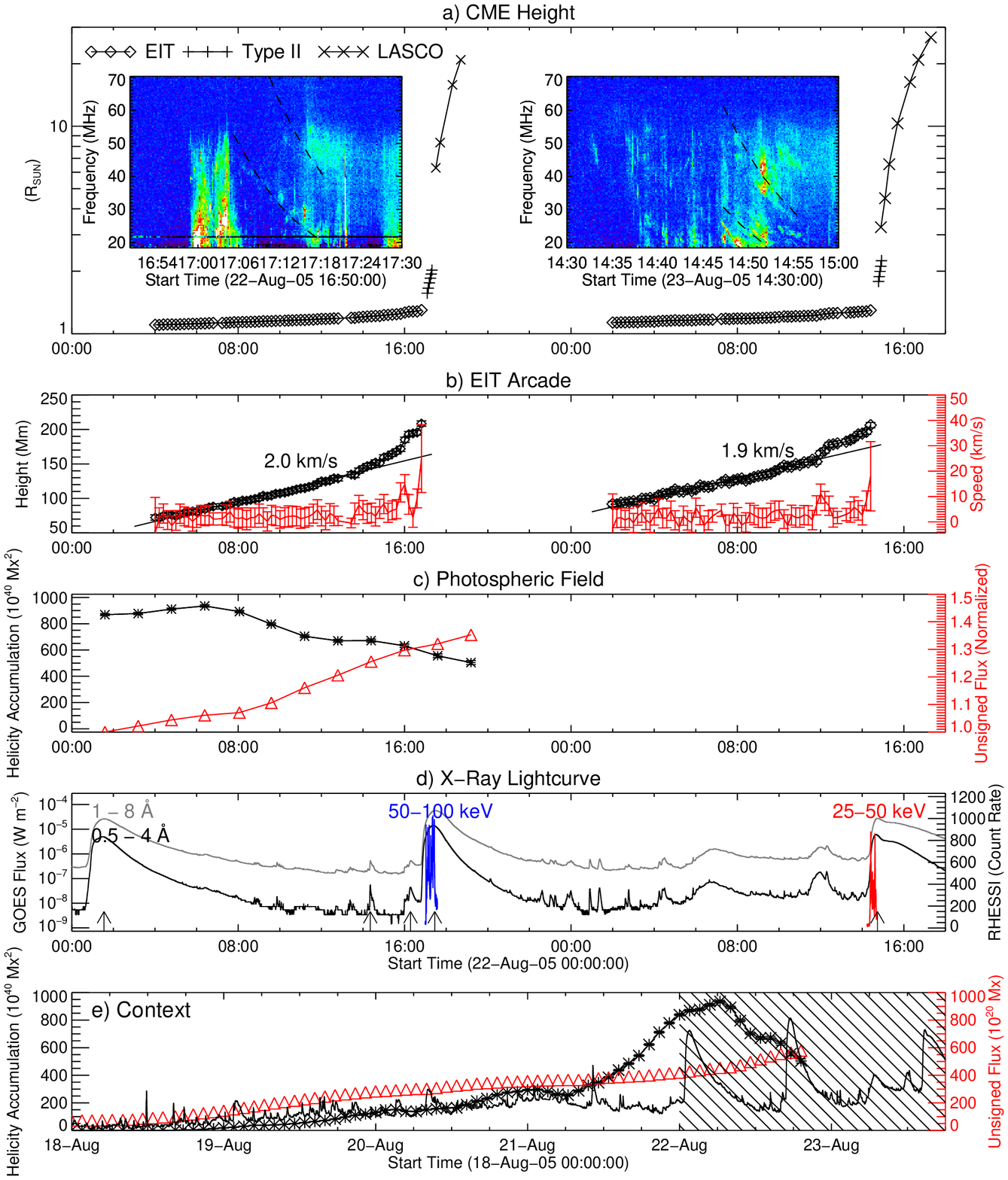} \caption{Height-time profile of the two PEAs and the resultant CMEs on 2005
August 22 and 23 in relation to the evolution of the photospheric magnetic field as well as X-ray
lightcurves. The insets of Panel (a) show the radio dynamical spectra recorded by the Green Bank
Solar Radio Burst Spectrometer (GBSRBS). For each flare of \sat{goes}-class C and above occurring
in AR 10798, we draw an arrow at the bottom of Panel (d) to indicate the soft X-ray flare peak.
Panel (e) show the helicity accumulation (asterisk) and flux emergence (triangle) of AR 10798 in a
larger temporal context than Panel (c), with the hatched box indicating the time duration covered
by Panels (a--d). \sat{goes} 1--8 {\AA} flux is displayed in an arbitrary unit.\label{ht0508}}
\end{figure}

The PEAs on August 22 and 23 were located in the same active region NOAA 10798, and their evolution
involved three homologous M-class flares associated with three Halo CMEs. The first PEA was
produced by an M2.6 flare (W54S11) and an accompanying Halo CME at about 01:30 UT on August 22. The
expansion and subsequent eruption of the arcade was associated with an M5.6 flare (W65S13) and the
second Halo CME (\fig{eit0508}(d)) at about 17:30 UT on August 22 (left column of \fig{eit0508}).
The eruption yielded the second PEA, whose expansion and subsequent eruption was associated with an
M2.7 flare on the limb (W90S14) and the third Halo CME (right column of \fig{eit0508}) at about
14:30 UT on August 23. This eruption created the third PEA, which, however, failed to erupt.

The difference between the first two PEAs that subsequently erupted is that, for the earlier
arcade, its eruption seems to involve the whole arcade of loops (\fig{eit0508}(a--c)), while for
the later one, the eruption involved only a group of loops which were apparently highly sheared
(marked by arrows in \fig{eit0508}(f) and (g)), with the loop-plane oriented in a similar
north-south direction as the neutral line. In addition, the gradual growth of the earlier arcade
started during the decay phase of the M2.6 flare till the subsequent eruption that occurred about
six hours after the end of the M2.6 flare; hence it is difficult to differentiate the growth due to
reconnections proceeding into higher altitudes, from the growth of the loops themselves. But for
the later arcade, the sheared loops only became distinctively visible when the rest of the arcade
became too diffuse to be seen at about 04:00 UT on August 23. At that time, the soft X-ray flux had
decayed to the background level; hence the contribution to the loop growth from reconnections can
be reasonably ignored. Nevertheless, both arcades exhibited quite similar kinematic characteristics
(\fig{ht0508}(a) and (b)). They both ascended slowly from $\sim$\rsun{1.1} to $\sim$\rsun{1.3} for
about 12 hours at \aspeed{2}. Only at the end of the quasi-static stage, for about tens of minutes,
the speed increased to tens of kilometers per second. Within one hour, both CMEs were accelerated
to \aspeed{2000}.

When inspecting the line-of-sight magnetograms, it came to our notice that significant flux
emergence was associated with the quasi-static expansion of the first PEA (see \fig{ht0508}(c)).
Meanwhile, one can see that the amount of positive helicity accumulation has continuously increased
for about 4 days (see \fig{ht0508}(e)), and only started to decrease during the gradual phase of
the first M-class flare, coincident with an enhanced rate of flux emergence. Hence the newly
emerged flux must be of opposite sense of helicity to the existing field. This has an important
implication for the pre-CME structure, and will be discussed in \S3.2. Note that the correction,
$B_n=B_l/\cos\psi$ (\S2.1), becomes increasingly inaccurate as the target region moves further away
from disk center. Hence we discontinued the calculation of the helicity injection when the center
of the active region went beyond 60 degree to the west of the central meridian (2005 August 22
19:11 UT).



\subsection{Eruption of Overlying Arcades}
\begin{figure}\epsscale{0.9}
\plotone{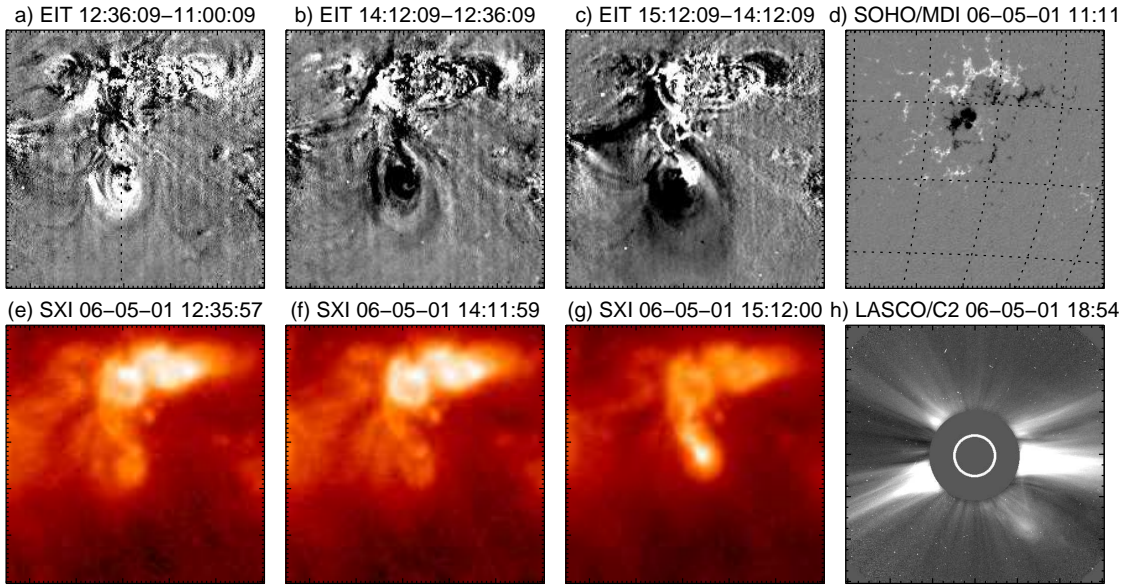} \caption{Evolution of the OA observed on 2006 May 1. The field of view in
Panels (a--g) is 550 by 550 arcsecs, centering at $(425'',\ -225'')$, with all images registered to
the image in Panel (a). \label{eit0605}}
\end{figure}

\begin{figure}\epsscale{0.9}
\plotone{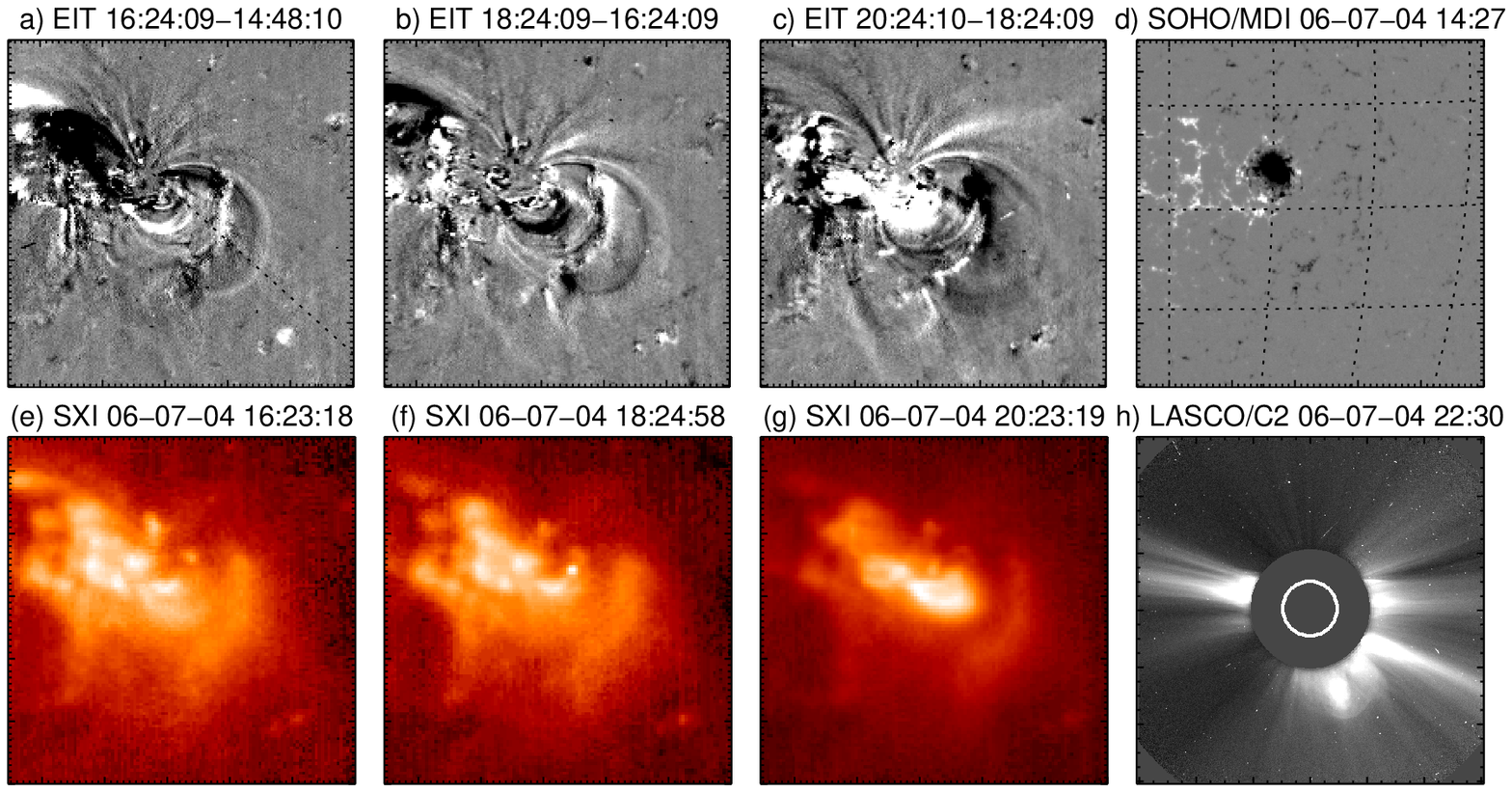} \caption{Evolution of the OA observed on 2006 July 4. The field of view in
Panels (a--g) is 550 by 550 arcsecs, centering at $(225'',\ -225'')$, with all images registered to
the image in Panel (a). \label{eit0607}}
\end{figure}

\begin{figure}\epsscale{0.9}
\plotone{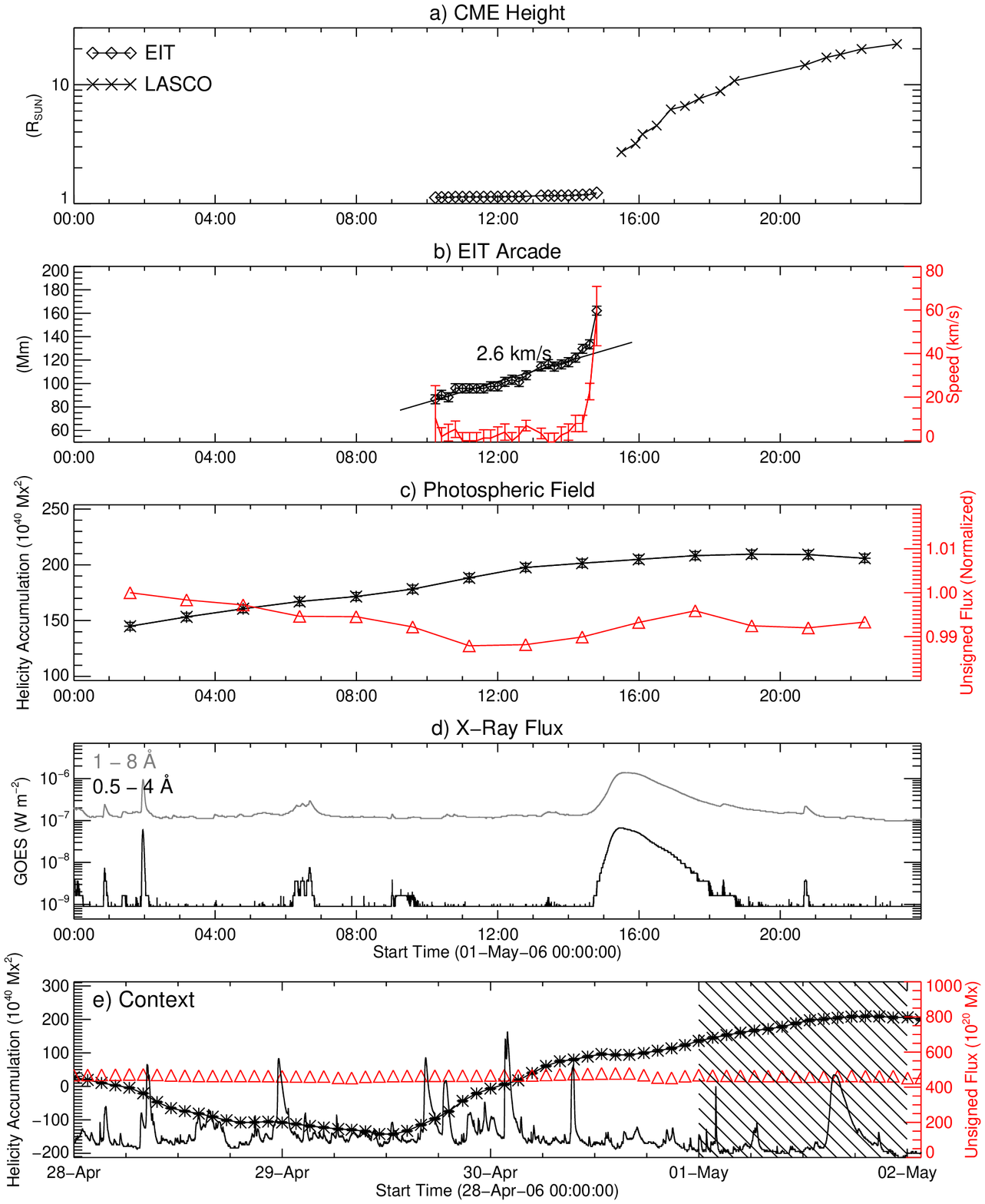} \caption{Height-time profile of the OA and the resultant CME on 2006 May 1
in relation to the evolution of the photospheric magnetic field as well as X-ray lightcurves. In
Panel (e), the hatched box indicates the time duration covered by Panels (a--d). \sat{goes} 1--8
{\AA} flux is displayed in an arbitrary unit. \label{ht0605}}
\end{figure}

\begin{figure}\epsscale{0.9}
\plotone{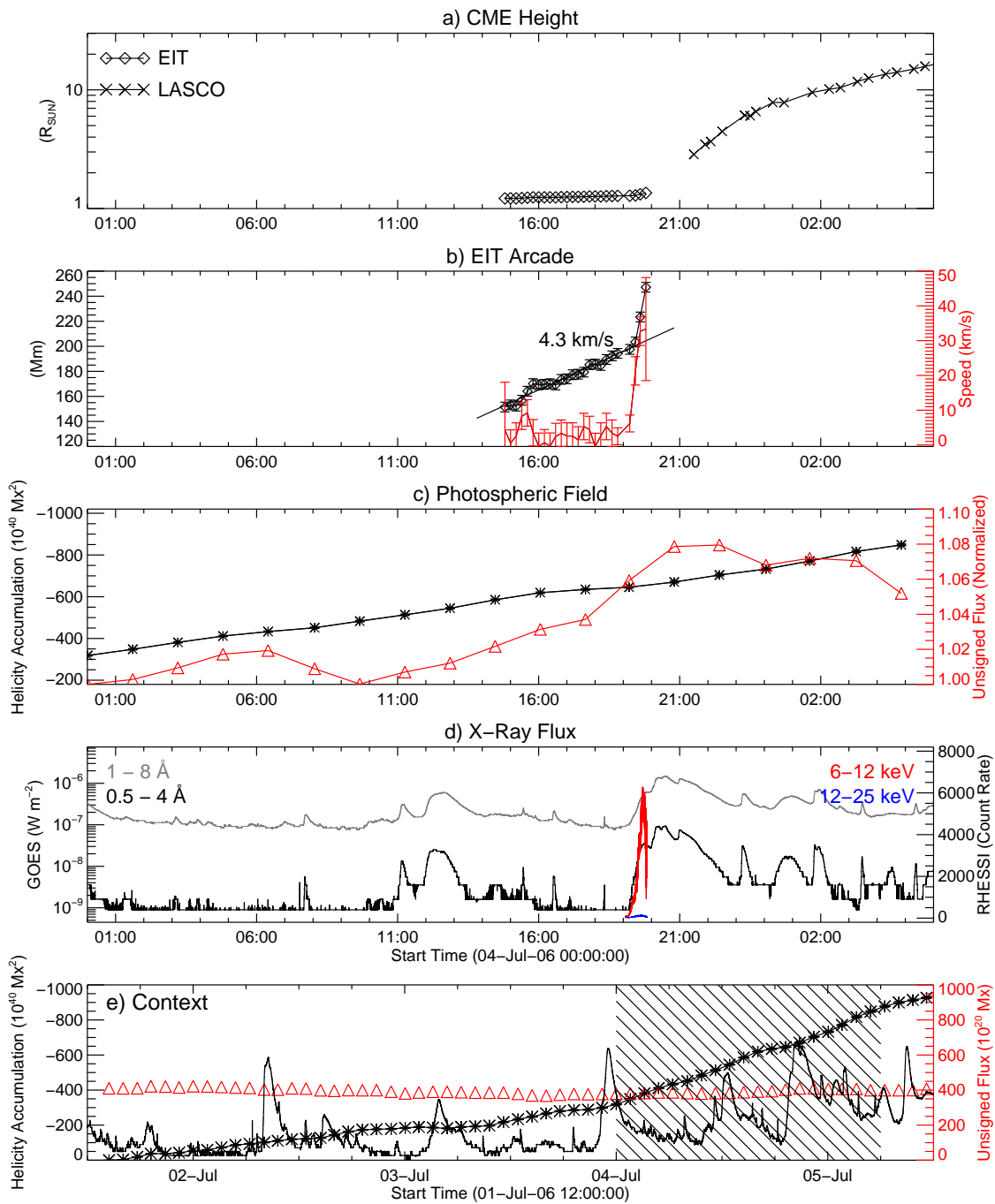} \caption{Height-time profile of the OA and the resultant CME on 2007 July 4
in relation to the evolution of the photospheric magnetic field as well as X-ray lightcurves. In
Panel (e), the hatched box indicates the time duration covered by Panels (a--d). \sat{goes} 1--8
{\AA} flux is displayed in an arbitrary unit. \label{ht0607}}
\end{figure}

The eruption of the two OAs on the disk, namely, Events 15 (\fig{eit0605} and \fig{ht0605}) and 16
(\fig{eit0607} and \fig{ht0607}), were observed on 2006 May 1 and 2006 July 4, respectively. With
the solar minimum impending, both events are much less energetic than the earlier ones. Putting
side by side, they exhibit quite similar characteristics. Both arcades were overlying a decayed
active region, namely, NOAA 10875 (\fig{eit0605}(d)) and NOAA 10898 (\fig{eit0607}(d)),
respectively. In both events, the gradual inflation of the OA is sustained for about 4.5 hrs, at a
speed of $<$\speed{5}, and the subsequent eruption only resulted in a C-class flare
(\fig{ht0605}(d) and \fig{ht0607}(d)). Both flares were associated with slow CMEs, although in the
former event a halo CME was produced (\fig{eit0605}(h)). Like the PEA events, the morphology of the
resultant CMEs bear similarities to the inflating arcades.

On the other hand, in the 2006 May 1 event, the arcade can be seen back to as early as 2006 April
28, 4 days prior to its eruption, while in the 2006 July 4 event, the overlying loops of interest
only became illuminated at 14:48 UT, and their growth and subsequent eruption were observed
henceforth. The loops on 2006 July 4 were located high in the corona from the beginning: the
projected half length of the highest loop is about \rsun{0.22}.

One may wonder how these bipolar, potential-like loops became eruptive and resulted in CMEs.
\sat{goes} soft X-ray images show highly complex loops underlying the inflating arcade in both
active regions (\fig{eit0605}(e--g) and \fig{eit0607}(e--g)), but there is no sign of twisted or
sheared fields, such as the well-known soft X-ray sigmoids. The quasi-static stage in both events
were temporally associated with helicity injection, except that the amount of helicity accumulation
flattened after the flare in the 2006 May 1 event (\fig{ht0605}(c)), while the helicity injection
rate displayed no obvious change throughout the flare in the 2006 July 4 event (\fig{ht0607}(c)).

\section{Discussion \& Conclusion}

\subsection{CME Initiation \& Kinematics}

\citet{zhang01} showed a three-phase kinematic evolution for three of four CMEs well observed by
LASCO C1, C2 and C3 coronagraphs from 1.1 to \rsun{30}: the initiation phase, impulsive
acceleration phase, and propagation phase. The initiation phase is characterized by a slow
ascension ($<$ \speed{80}) over tens of minutes, which always occurs before the onset of the
associated flare. The impulsive acceleration phase with a rapid acceleration of 100--500 m
s${}^{-2}$ coincides very well with the flares's rise phase lasting for a few to tens of minutes.
The acceleration of CMEs ceases near the peak time of the soft X-ray flares. The final phase is a
propagation at a constant or slowly decreasing speed. This temporal correlation between the CME
velocity and the soft X-ray flux of the flare is further confirmed in CMEs characterized by
intermediate and gradual acceleration \citep{zhang04}.

CMEs in our observations, however, feature a quasi-static inflation phase of the coronal arcade at
$<$\speed{5} for about 4--12 hrs, followed apparently by a similar three-phase paradigm as
established by \citet{zhang01,zhang04}, during which the arcade evolves into the CME front. The
gradual inflation of both PEAs and OAs seems to be a response of coronal magnetic fields to the
continued injection of magnetic free energy from below, via flux emergence, or photospheric flows,
as demonstrated by the increasing helicity accumulation prior to the eruption. One may argue that
for PEA events the preceding eruption might not release all of the free energy available, which
makes the subsequent eruption possible. However, since the quasi-static stage lasts for hours,
during which the PEA sometimes survives multiple flares, we suppose that the arcade is quite
stable, otherwise, a little additional energy supply or disturbance might have triggered its
eruption. Therefore, it is reasonable to assume that a significant portion of the energy powering
the eruption of the PEA is accumulated during the quasi-static stage. In general, the timescale of
the quasi-static stage is dependent on both the stability of the pre-CME structure and the
accumulation rate of magnetic free energy in the corona. There may exist a distribution of the
timescale spanning from hours (for the events studied in this paper) to days (for helmet
streamers). For example, \citet{sw07} observed the gradual inflation of much higher coronal loops
in the LASCO FOV, which sustains for 1--2 days at \aspeed{20} and ends with the sudden formation of
a pair of inward and outward components moving at speeds of \aspeed{100} and \aspeed{300},
respectively. Nevertheless, we suppose that a quasi-static stage, which corresponds to the energy
accumulation in the corona, is inherent to the kinetic evolution of any CME, no matter if the
pre-CME structure has the right temperature and density to be seen in a narrow filter like EIT 195
{\AA}.

\subsection{Pre-CME Configuration}
\begin{figure}\epsscale{0.6}
\plotone{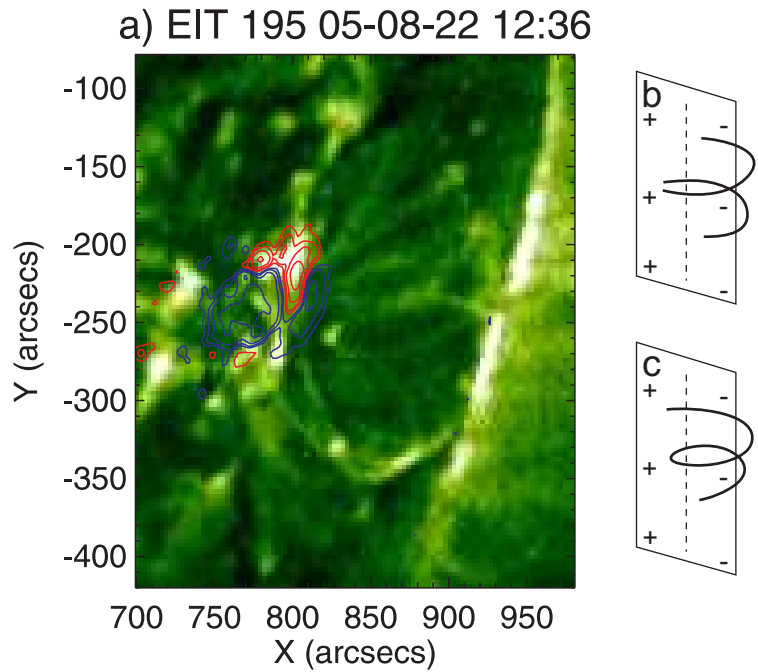} \caption{Swirling structure observed in the center of the 2005 August 22
arcade. In Panel (a) overlaid above the EIT 195 {\AA} image at 12:36 UT are contours of an MDI
magetogram taken at approximately the same time. Contour levels are 100, 200, 400 and 800 G for
positive polarities (red), and -800, -400, -200, and -100 G for negative polarities (blue). Panels
(b) and (c) indicate two alternative interpretations of the swirling structure. \label{fluxrope}}
\end{figure}

While it is generally agreed that the free energy powering CMEs is most likely stored in stressed
(twisted or sheared) fields, there has been contentious debate over the nature of the pre-eruption
configuration. The debate focuses on two competing models, namely, flux rope models vs.
sheared-arcade models. In the sheared-arcade models, a flux rope is formed via magnetic
reconnection during the course of the eruption \citep[e.g.,][]{adk99}. Hence it has been argued
that a pre-existent flux rope is unnecessary for solar eruptions. On the other hand, the flux rope
configuration provides a natural explanation for the three-part structure of CMEs and their
quiescent counterparts, i.e., helmet streamers \citep{low96, low01}. CMEs that exhibit circular
intensity patterns have also been interpreted as a manifestation of helical magnetic fields,
therefore being termed flux-rope CMEs \citep[e.g.,][]{dere99}. Recently, \citet{gm04} reported the
eruption of a multi-turn helix from within a region of sheared magnetic field. The presence of
twisting and kinking motions in eruptive prominences further argues for the existence of flux ropes
in the corona \citep[e.g.,][]{alg06,lag07}. While these morphological studies demonstrate that flux
ropes are indeed associated with CMEs, it is still unknown whether they are present prior to solar
eruptions, although their presence ``after'' the eruption have been confirmed by \emph{in-situ}
observations of rotating magnetic fields \citep{burlaga81}.

The inflation of coronal fields can be attributed either to the shearing of the magnetic
footpoints, or to the emergence of new flux. However, the eruption of the whole PEA such as the
2005 August 22 event (\fig{eit0508}(a--c)) poses a severe constraint on the energetics of the
eruption, since it has been demonstrated that a bipolar force-free field is at its maximum energy
when the field is completely ``open'' \citep{aly84, aly91, sturrock91}. A multipolar topology as
proposed in the break-out model \citep{adk99} can help to circumvent the so-called Aly-Sturrock
limit, since only one of the bipolar arcades is opened up. Indeed AR 10798 that hosted the 2005
August 22 PEA displays a quadrupolar field topology, as demonstrated by the contours of the
\sat{soho} MDI magnetogram (\fig{fluxrope}(a)).

Alternatively, studies have shown that the Aly-Sturrock limit can be bypassed if the coronal field
contains a detached flux rope \citep[e.g.,][]{wolfson03,flyer04}. Intriguingly, a swirling
structure can be seen in the center of the 2005 August 22 arcade as early as 12:12 UT
(\fig{eit0508}(b)), and was most obvious at 12:36 UT (\fig{fluxrope}(a)), over four hours before
the onset of the M5.6 flare at about 16:46 UT. Ambiguity in interpreting this structure exists due
to the limitation of the two-dimensional observation. It could be composed of two sheared loops as
illustrated in \fig{fluxrope}(b), or a truly twisted structure as in \fig{fluxrope}(c). The former
configuration, however, requires the two loops to be sheared in opposite directions on both sides
of the neutral line (denoted by the dashed line). Hence, the latter configuration, namely, a flux
rope, offers a more natural explanation. For the 2005 January 15 event (\S2.2.1), \citet{cheng10}
showed that a flux rope was located below the rising arcade via nonlinear force-free-field
modeling.

As part of the PEA, the flux rope must be generated in the corona, via three possible ways, viz.,
a) reconnection of sheared magnetic fields \citep[e.g.,][]{vm89}; b) reconnection within the flux
rope involved in the preceding eruption, which is therefore a partial eruption \citep{gal07}; and
c) the emergence of a fresh magnetic field of the opposite helicity into a preexisting coronal
field, which is observed for the 2005 August 22 event (\S2.2.2). \citet{zl03} argued that with the
flux emergence reconnection should take place between the two flux systems to take the field to a
minimum-energy state, and that this relaxation process that keeps the total helicity conserved may
result in the formation of magnetic flux ropes. Thus, the pre-CME structure like the 2005 August 22
PEA may posses a pre-existent flux rope, a magnetic configuration that many suggest for the cavity
structure of the helmet streamer. We have noticed that helmet streamers undergo similar slow
inflation before erupting into interplanetary space \citep[e.g.,][]{gibson06}, and that some
coronal helmet streamers were reported to be temporarily visible in X-rays, high above the rising
post-flare loops \citep[e.g.,][]{svestka97}.

\subsection{Flare-CME Relationship}

There is a longstanding debate on the flare-CME relationship in the solar physics community:
whether flares are the cause of CMEs or the other way around \citep{gosling93}. \citet{harrison95}
presented a comprehensive review on the flare-CME relationship before the launch of \sat{soho}, and
concluded that they do not drive one another but are closely related. Using \sat{soho} LASCO data,
\citet{zhang01, zhang04} demonstrated that a close temporal correlation exists between the CME
velocity and the soft X-ray flux of the flare. Due to the Neupert effect \citep{neupert68}, a
similar correlation exists between the CME acceleration and the derivative of soft X-ray flux,
suggesting that the CME large-scale acceleration and the flare particle acceleration are strongly
coupled. \cite{lin04} treated the flare, prominence and CME as integral constituents of a single
process within the framework of the catastrophe model, and suggested that the flare-CME correlation
depends on the free energy stored in the relevant magnetic structure: the more free energy, the
better correlation.

\citet{zl05}, alternatively, suggested that flares and CMEs play different roles in the MHD
processes driving eruptions. They noticed that although flares can dissipate excessive magnetic
free energy, it is CMEs that shed the excessive helicity in the corona. Due to the probable
existence of an upper bound on the total magnetic helicity in the corona \citep{zfl06}, CMEs could
be the consequence of accumulating helicity which is generated by the dynamo and transported
through the photosphere into the corona. This physical view is supported by some statistical
studies, e.g., \citet{na04} found that, in a statistical sense, active regions producing flares
associated with CMEs have a larger quantity of estimated magnetic helicity than those producing
flares without any CME.

Our observations show some characteristics in support of \citet{zl05}. The quasi-static inflation
of the pre-CME arcade sustains for hours, during which multiple flares have usually occurred in the
same active region, but fail to affect the evolution of the arcade. Thus, the pre-CME structure,
namely, the inflating arcade, is arguably independent of the flares during the quasi-static phase,
but is closely coupled with the flare during the acceleration phase. Taken Event 6 for example
(\fig{ht0501}), at least 11 flares, including two M-class flares and nine C-class flares, occurred
in the vicinity of the arcade during its quasi-static inflation stage, but none of them has
significant effect on the evolution of the arcade. Are flares with and without CMEs distinctly
different from each other? \citet{yashiro06} reported that the power-law distributions for peak
fluxes, fluences and duration are significantly steeper for flares without CMEs than for flares
associated with CMEs. Further investigation is desired on the underlying physics leading to the
distinctive statistical characteristics.

To summarize, we have identified a group of active-region coronal arcades, which mainly consist of
post-eruptive arcades, and whose gradual inflation build up to CMEs. The quasi-static inflation
stage sustains for hours at a speed of less than \speed{5}. It is temporally associated with
significant helicity injection from photosphere, and followed by a frequently observed three-phase
CME evolution paradigm, as the arcade, which may have survived multiple flares, suddenly erupts as
a CME.

\acknowledgments SOHO is a project of international cooperation between ESA and NASA. The authors
were supported by NASA grants NNX08-AJ23G, NNX08-AQ90G, and NNX08-BA22G, and by NSF grant
ATM-0849453 and ATM-0819662.


\end{document}